\documentclass[twocolumn,showpacs,amsmath,amssymb,prb,graphics,graphicx]
{revtex4-2}
     \usepackage{graphicx}
\usepackage{bm}
\usepackage{dcolumn}
\begin{document}

\title{Vortex in   superconducting  thin-film strips of arbitrary width}
\author{N. Nakagawa}
\address{Iowa State University, Ames Iowa 50011, USA }
\author{V. G. Kogan}
\address{Ames Laboratory - Department of Energy, Ames Iowa 50011, USA }

\begin{abstract}
The currents and field distributions of a vortex in a thin 
superconducting strip of a width $W$ is considered.  
It is shown that unlike infinite films where the vortex field crosses the film only in one direction (say, from the half-space under the film to the half-space above it), in strips (as well as in finite size film samples of any shape) the field lines 
go back to the lower half-space not only out of the sample but also through certain strip parts. The strip patches where the magnetic lines ``dive" back to the space under the strip are situated mostly near the strip edges at the strip and out of it. 
 The magnetic flux through the strip associated with the vortex is shown to be less than the flux quantum and depends on the vortex position.  The suppression of the vortex flux is strong 
 in narrow strips with $W<<\Lambda=2\lambda^2/d$ where $\lambda$ is the bulk London penetration depth and $d$  is the film thickness. 
  The   vortex energy   scales roughly with $W/\Lambda$ and depends on vortex position.
\end{abstract}
%\pacG{PACG {\bm 74.76.-w}}
\date{\today}
 \maketitle

\section{Introduction}

Thin film superconducting strips are ubiquitous elements  in superconducting electronics. In particular, they are used in  Single Photon Superconducting Detectors, which are basically made of these strips fed with a  transport current close to critical when the condensation energy is small and even a single photon can cause transition to the normal state, recorded as a voltage pulse. In this set up, however, vortices can nucleate at the strip edges and cross the strip also producing dissipation, see e.g. \cite{Schilling,Karl} and references therein. To study these so-called dark counts, one has to know properties of vortices in thin film strips, called Pearl vortices after J.Pearl who studied them in infinite films \cite{Pearl}.

Unlike the field of Abrikosov vortex in the bulk, localized within the area of a size $\lambda$ of the London penetration depth, the large part of the Pearl vortex energy is stored in the stray field outside the film that decays as a power law. Pearl vortices with their stray fields are in fact three-dimensional (3D) entities. The stray fields distributions in vacuum are not bound either by the London   $\lambda$ or by the Pearl length $\Lambda=2 \lambda^2/d$, but rather determined by persistent current distributions and the film sample shape. Hence, the structure of vortices in thin-film strips may differ substantially from the case of infinite films. To our knowledge this question - at first sight rather academic - has not been addressed in literature. In this publication we attempt to fill this ``white spot". 

We argue that for narrow strips of  a width $W\ll \Lambda$, the self-field of the vortex  can be disregarded, so that the current distribution can be calculated first after which the field evaluation is straightforward. For the general case of arbitrary widths
both currents and fields should be found simultaneously by solving an integral equation, a more involved procedure. Also, we calculate the magnetic flux through the strip and the vortex energy as functions of the vortex position.

 Our approach is based on  the London equation averaged over the film thickness $d$,  
\begin{equation}
h_z+2\pi\Lambda \, {\rm curl}_z {\bm g}/c=\phi_0 \delta ({\bm r}-{\bm a})\,, 
\label{e1}
\end{equation}
where ${\bm g}({\bm r})$ is the sheet current density, ${\bm r}=(x,y)$, ${\bm a}=(a,0)$ is the vortex position, and 
$\Lambda =2\lambda^2/d$ is the Pearl length.  Equation (\ref{e1}) is valid everywhere at the film except the vortex core and a narrow belt  of a width $\xi$ adjacent to the edges, where the London equations break down  \cite{Larkin}. 

Currents ${\bm g}({\bm r})$ and the field $h_z({\bm r})$ at the film plane $z=0$ can be found by solving Eq.\.(\ref{e1}) 
combined with the continuity equation and the Biot-Savart integral which 
relates the field $h_z$  to the surface current:
\begin{equation}
{\rm div} {\bm g}=0\,,\,\,\,\,\, 
h_z({\bm r})c=\int d^2 {\bm r}'[{\bm g}({\bm r}')\times {\bm R}/R^3]_z\,; 
\label{e2}
\end{equation}
${\bm R}={\bm  r}-{\bm  r}'$. 
The specific feature of the thin film limit should be noted: since the derivatives $\partial/\partial z$ in the film are 
large relative to  $\partial/\partial {\bm r}$, the Maxwell equation ${\rm curl} {\bm h}=4\pi {\bm j}/c$ 
is reduced to conditions relating the sheet current to discontinuities of 
the tangential field \cite{LL}: 
\begin{equation}
2\pi g_x/c=-h_y(+0),\,\,\, 2\pi g_y/c=h_x(+0)\,. 
\label{e3}
\end{equation}
Here, $h_{x,y}(+0)=-h_{x,y}(-0)$, and $\pm 0$ stand for the upper and lower faces of the 
film. The field  $h_z$  is related to currents by an integral (\ref{e2}), rather than by a differential equation.  

Equations (\ref{e1}) and ({\ref{e2}) along with Maxwell equations for the stray field form a complete set for determination of currents and fields. Instead of $\bm g$, it is 
convenient to deal with a scalar stream function $S({\bm r})$ such that
${\bm g}={\rm curl}S{\hat {\bm z}}$:
\begin{equation}
 g_x=\partial_yS \,,\,\,\, g_y=-\partial_xS\,. 
\label{e4}
\end{equation}
Then the first of Eqs.\,(\ref{e2}) is satisfied. 

The kernel ${\bf R}/R^3$ of the Biot-Savart integral is strongly singular.
To reduce the degree of singularity one can 
write ${{\bf R}/R^3}=\nabla '(1/R)$ (the prime 
specifies ${\bf r}'$ as the variable of differentiation) and 
integrate by parts. Equation (\ref{e2}) then becomes
\FL
\begin{eqnarray}
&&h_zc=\int_{\rm strip} \frac{d^2 {\bm r}'}{R}{\rm curl}_z{\bf g}({\bm r}') \nonumber\\
&&+ \int_{-\infty}^{\infty}dy' \frac{g_y({\bm r}')}{R}\Big |_{x^\prime=0} +
\int_{ \infty}^{-\infty}dy' \frac{g_y({\bm r}')}{R}\Big |_{x^\prime=W}\,,\qquad
\label{e5}
\end{eqnarray}
where $W$ is the strip width.   Eq.\,(\ref{e1}) can now be written as an equation for $S(\bm r)$:
          \begin{eqnarray}
&&\int_{\rm strip}d^2{\bm r'}\frac{\nabla^2S({\bm r}')}{R} + \int_{-\infty}^
{\infty}dy'\frac{\partial_{x'}S({\bm r}')}{R}\Big |_{x^\prime=0} \nonumber\\
&&+ \int_{\infty}^
{-\infty}dy'\frac{\partial_{x'}S({\bm r}')}{R}\Big |_{x^\prime=W}   
+ 2\pi\Lambda\nabla^2S({\bm r})=-c\phi_0 \delta({\bf r}-{\bf a}).\nonumber\\    
\label{e6}
          \end{eqnarray}
This is to be solved for $S(x,y)$ at the strip $0<x<W$ subject
to boundary conditions ${\bf g(\infty)}=0$ and the vanishing
normal component of the current at the film edges $g_x(0,y)=g_x(W,y)=0$. 
These conditions imply a constant $S$   at
the film edges; one can set $S=0$ at the film boundaries
since only the derivatives of $S$ have physical meaning.

%%%%%%%
\section{Narrow strips, $\bm{W<<\Lambda}$}
%%%%%%

The  integro-differential Eq.\,(\ref{e6}) holds for any $W$ and in general can be dealt with   numerically. The situation simplifies for $W\ll \Lambda$:
as is seen from the Biot-Savart integral (\ref{e2})  $h_z \sim g/c$, whereas $2\pi\Lambda \, {\rm curl}_z {\bm g}/c \sim (\Lambda/W)g/c\gg h_z$. Hence, $h_z$ in the basic Eq.\,(\ref{e1}) can be disregarded  for narrow strips, so that one can first  solve the truncated Eq.\,(\ref{e6}),
          \begin{eqnarray}   
  2\pi\Lambda\nabla^2S({\bm r})=-c\phi_0 \delta({\bf r}-{\bf a}),    
\label{e7}
          \end{eqnarray}
for  $S({\bm r})$ and obtain  currents. Then one can use the converted  Biot-Savart  Eq.\,(\ref{e5}) to evaluate the field. 

The problem of solving for $S$  is 
equivalent to one in 2D electrostatics with a known solution \cite{Smithe,Morse,BGBK,K94}:
\begin{equation}
S(\bm r)=\frac{c\phi_0}{8\pi^2\Lambda} 
 \ln\frac{\cosh \pi y-\cos \pi (x+a) }{\cosh \pi y -\cos \pi(x-a) }\,,
\label{S1}
\end{equation}
where we use $W $ under the log-sign  as  unit length (in these units $0<x<1$ at the strip). According to Eq.\,(\ref{e4}) the currents  in units of $c\phi_0/4\pi \Lambda W$ are
          \begin{eqnarray}   
  g_x=  \frac{\sin\pi a \sin\pi x \sinh \pi y }{[\cos \pi(x-a)-\cosh \pi y ][\cosh \pi y -\cos \pi(x+a)] },\qquad \nonumber\\
  g_y=  \frac{\sin\pi a (\cos\pi a - \cos\pi x \cosh \pi y }{[\cosh \pi y -\cos \pi(x-a) ][\cosh \pi y -\cos \pi(x+a)]}.\qquad   
\label{currents}
          \end{eqnarray}
 The current flow lines are   contours $S=$\,const; an example is shown in Fig.\,\ref{f1}.
  \begin{figure}[h ]
\includegraphics[width=7cm] {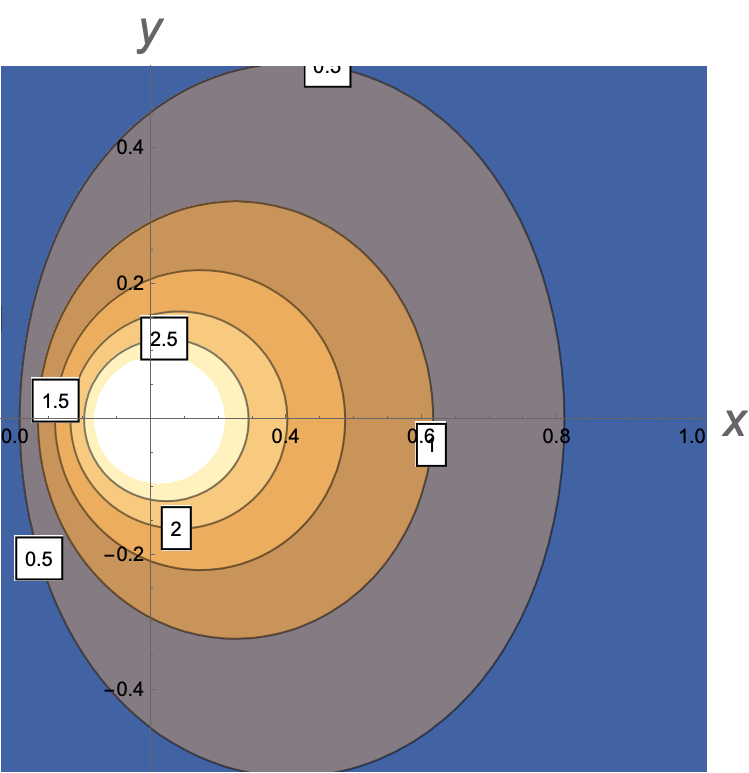}
\caption{Contours of constant $S(x,y)$ in units  $c\phi_0/8\pi^2\Lambda$  (or the current streamlines) for a vortex at $a=0.2 W$. $x,y$   are in units $W$ so that the strip edges are  $x=0$ and $x=1$.   }
\label{f1}
\end{figure}

Given the currents, Eq.\,(\ref{currents}), the Bio-Savart law of Eq.\,(\ref{e2}) provides $h_z(\bm r)$ everywhere at the plane $z=0$. The calculation implies the double integral over the strip, a heavy numerical task done with the help of the Mathematica package. 
Figure \ref{f2} shows $h_z(x,0)$ for $a=0.2$   in  units of $W$. 
  \begin{figure}[h]
\includegraphics[width=8cm] {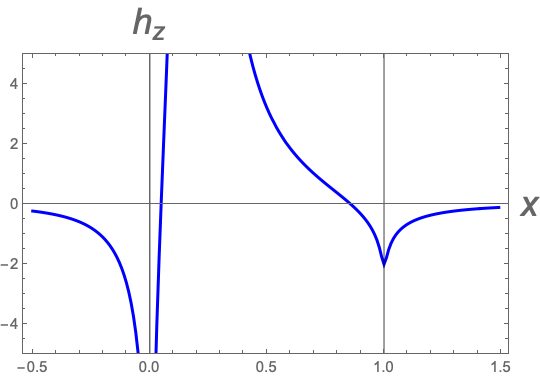}
\caption{The field $h_z$ vs $x$ for $y=0$, $z=0.01$,   and  $a=0.2$.  $x, z $, and $a$ are in units $W$ so that the strip edges are at $x=0, 1$.   }
\label{f2}
\end{figure}
Without going to formal details, we note that $h_z(x,0)$ goes as $1/|x-a|$   near the vortex core. 
 
The regions on the left and right of the strip of negative $h_z$  are due to the flux going round the edges from the   space above the film to one under it.       
The most interesting, however, is that the sign change of $h_z$ happens not at the edges {\it per se}, but in their vicinity {\it at the strip}. In the example of Fig.\,\ref{f2}, the sign change positions are  at $x\approx 0.05$ and $x\approx 0.8$ at $y=0$. The region of positive $h_z$ is strongly elongated along the strip;   for $a=0.2$, this region extends up to $|y|\approx 1.1 $. 

Now we recall that  in infinite {\it isotropic} films, $h_z>0$ everywhere \cite{Pearl}. However, it was shown in \cite{KNK} that $h_z$ can change  sign at some parts of  {\it infinite anisotropic}  thin films. We see now that this may happen also in isotropic thin fim strips. The domain of negative $h_z$ is clearly seen in the 2D contour plot of Fig.\,\ref{f3}. The white patches of this plot correspond to $|h_z(x,y)|>1$. In fact,  $h_z\to \infty$  at the vortex position $(0.2,0)$. The  divergence  of the type $1/r$ is an artifact  of the London theory and in reality is truncated at distances of the order of coherence length $\xi$.
  \begin{figure}[htb ]
\includegraphics[width= 8cm] {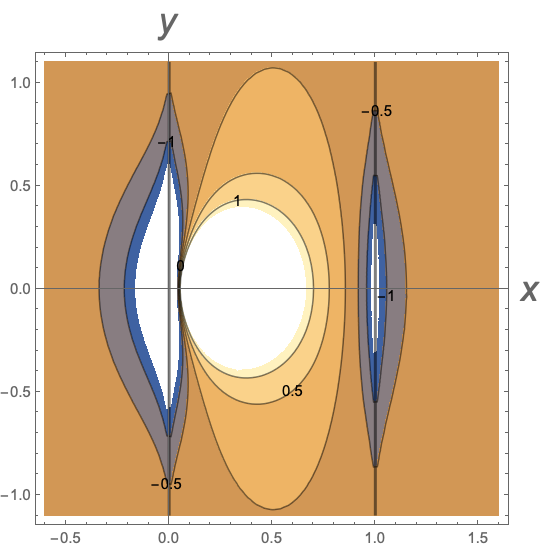}
\caption{Contours of the constant field $h_z(x,y)$ in units of $\phi_0/4\pi \Lambda W$ for the vortex at $\bm a=(0.2,0)$.   The solid vertical lines are the strip edges.   }
\label{f3}
\end{figure}

One may say that the major ``flux sinks" of dark colors in Fig.\,\ref{f3}  are situated along the edges within the strip and outside it. It is reasonable to expect that such patches exist when vortices reside in  finite film samples of any shape.
 
For the sake of completeness, we mention here other known results for narrow strips:

\subsubsection{Vortex self-energy} 

 In zero applied field and with zero transport current in narrow strips, the vortex energy  (of the stray field outside the film and the kinetic energy of persistent currents in the film) has been evaluated in \cite{K94,Clem,Maximova}:
\begin{equation}
\epsilon(a)=\frac{\phi_0^2}{8\pi^2\Lambda}\,{\rm ln}\left(
\frac{2W}{\pi\xi}\,{\rm sin}\frac{\pi a}{W}\right )\,,
\label{energy}
\end{equation}
where $\xi$ is the estimate of the core size (the coherence length).
In the presence of transport current, which is uniform in narrow strips, this energy has been discussed in \cite{Tafuri}.

\subsubsection{Interaction of vortices}
 
 For two vortices at $\bm a_1=(x_1,y_1)$ and $\bm a_2=(x_2,y_2)$, the interaction energy is \cite{BKT}:
\begin{equation}
\epsilon_{int}= \frac{\phi_0^2}{8\pi^2\Lambda} \ln\frac{\cosh (y_1-y_2) -\cos (x_1+x_2) } 
{ \cosh (y_1-y_2) -\cos (x_1-x_2) }\,,
\label{interaction}
\end{equation}
where the coordinates are given in units of $W/\pi$. 
It is easy to see that this interaction  exponentially decays for vortices situated at different $y$'s with the decay length of the order $W$ \cite{BKT}. 
It is worth noting that, unlike in the bulk case, the material parameter $\Lambda$ in narrow thin-film strips  enters $\epsilon_{int}$ only in the pre-factor $\phi_0^2/8\pi^2\Lambda$, whereas the coordinate dependence of interaction does not contain $\Lambda$ at all.

 %%%%%% 
\subsubsection{Flux through  the strip} 
%%%%%%%

It has been shown in \cite{K94} that this flux is
\begin{equation}
\phi_z(a)=\frac{\phi_0}{\pi\Lambda }\left (a\,{\rm 
ln}\frac{W-a}{a}+ W\,{\rm ln}\frac{W}{W-a}\right)\,.
\label{phi_z}
\end{equation}
The flux $\phi_z(a)$ turns zero at the edges (as $-a\,{\rm ln}a$ at 
$a\rightarrow 0$) 
and reaches maximum of $\phi_0W\ln 2/\pi\Lambda )$ 
in the strip middle. Thus, the flux 
carried by a vortex in a narrow ($W \ll \Lambda $) thin-film bridge 
scales with the ratio $W/\Lambda $, depends on the vortex position, and is 
much smaller than the flux quantum.

Similar to $\phi_z$, one can estimate the flux which goes around the film edges 
from the half-space above the strip to the lower half-space. One obtains for the flux 
$\phi_z^L=\int_{x<0}h_zd^2{\bf r}$ crossing the plane $z=0$ 
left of the edge $x=0$:  
\begin{equation}
\phi_z^L= -\frac{\phi_0 a}{ \pi\Lambda }\,{\rm ln}\frac{W}{a}\,.
\label{phi_Left}
\end{equation}
 The flux $\phi_z^L$ drops 
fast when the vortex approaches the left edge ($\phi_z^L\propto 
a\,{\rm ln}a$), whereas the decrease is slow for the vortex 
moving toward the opposite edge: $\phi_z^L\propto (W-a)$. The flux 
$\phi_z^R=\int_{x>W}h_zd^2{\bf r}$ at the right of $x=W$ 
is evaluated in a similar manner to show that 
$\phi_z^L+\phi_z^R+\phi_z=0$. Moreover, one can show that the total flux 
crossing any plane $z=const$ vanishes, unlike the case 
of a vortex in an infinite film where it is $\phi_0$. 

%%%%%%%%%%%%% 
\section{Arbitrary width}
%%%%%%%

For the general width $W$, we retain the $h_z$ term in Eq.\,(\ref{e1}), and then solve the coupled equations Eqs.\,(\ref{e1}) and (\ref{e2}) numerically.  For the strip geometry, we expand $S(\bm r)$ as a series of  eigenfunctions   of the 2D Laplacian   
\begin{equation}
	\chi_{n}^{k_y}(\textbf{r}) = e^{ik_yy} \sqrt{\frac{2}{W}}\sin \frac{\pi n x}{W},\ 0 \leq x \leq W,\; n=1,2,\cdots,
	\label{eNN1}
\end{equation}
that vanish  at the edges.  
 In particular, the narrow strip solution (\ref{S1}) is expandable as
\begin{align}
 S_ {\rm narrow} (\bm r) & = \frac{c\phi_0}{2\pi\Lambda}\sum_{n}\int_{-\infty}^\infty\frac{dk_y}{2\pi}\frac{\chi_n^{k_y}(\bm r) \chi_n^{k_y}(\bm a )} { ( n\pi/W )^2+k_y^2 } \nonumber
 \\
 & \equiv \frac{c\phi_0}{2\pi\Lambda} \overline{G}(\bm r ,\bm a ),
 \label{eNN2}
\end{align}
where $\overline{G}(\bm r ,\bm r' )$ acts as the Green's function of the strip problem, see Eq.\,(\ref{e7}). 
 
 The term $h_z$ in (\ref{e2}) can be considered as  extra source, yielding the relationship
\begin{equation}
	S(\bm r) = \frac{c}{2\pi\Lambda} \Big[ \phi_0\overline{G}({\bm r},{\bm a})
	- \int_{\rm strip} d\bm r' \overline{G} ({\bm r},{\bm r}^\prime ) h_z({\bm r}^\prime) \Big],\qquad\qquad
	\label{eNN3}
\end{equation}
which can be coupled with Eq.\,(\ref{e2}) to solve for $S$ after eliminating $h_z$.  Unlike $\overline{G}(\bm r ,\bm r' )$, the integration kernels of Eq.\,(\ref{e2}), i.e. $\bm{R}/R^3$, have off-diagonal elements with respect to the $\chi$ basis.  Hence, our task is to solve the coupled equations in the matrix form by means of truncation.  Numerical results can be obtained for arbitrary sets of parameters $W/\Lambda$ and $a/W$.  The actual numerical process requires to work around the divergence of the solutions at the vortex position, as outlined in the Appendix in some detail.

Typical stream functions $S(x,y)$  obtained numerically  are shown in Fig. \ref{fig4}.     
 It is worth noting that for $W/\Lambda <1$ the narrow strip solution (the left panel of the figure) does not change much. Since this is the case for strips commonly used in practice, the narrow strip description may often suffice.  
 
\begin{widetext}

\begin{figure}[h]
	\centering
		\includegraphics[width=1.\textwidth]{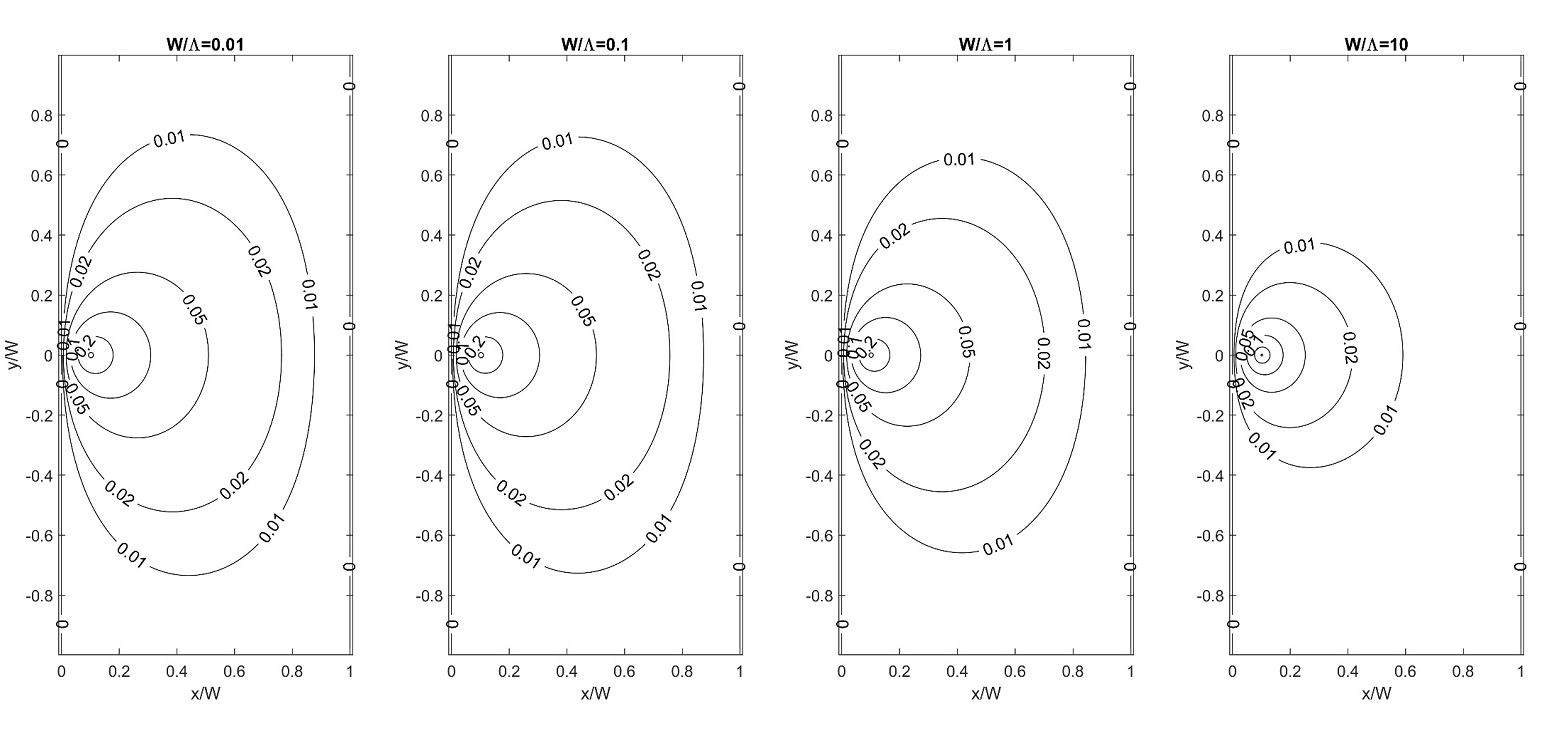}
	\caption{The stream function $S$  in the unit of $c\phi_0/2\pi\Lambda$ is plotted for   widths $W/\Lambda=\{0.01, 0.1, 1, 10\}$  for  the same  vortex position $a/W=0.1$.}
	\label{fig4}
\end{figure}

 This trend is reflected also in the calculated $h_z(\bm r)$.  Figures \ref{f5}  show examples, exhibiting resemblance in the field patterns for $W/\Lambda<1$ for the same vortex position.
\begin{figure}[h]
	\centering
		\includegraphics[width=1.	\textwidth]{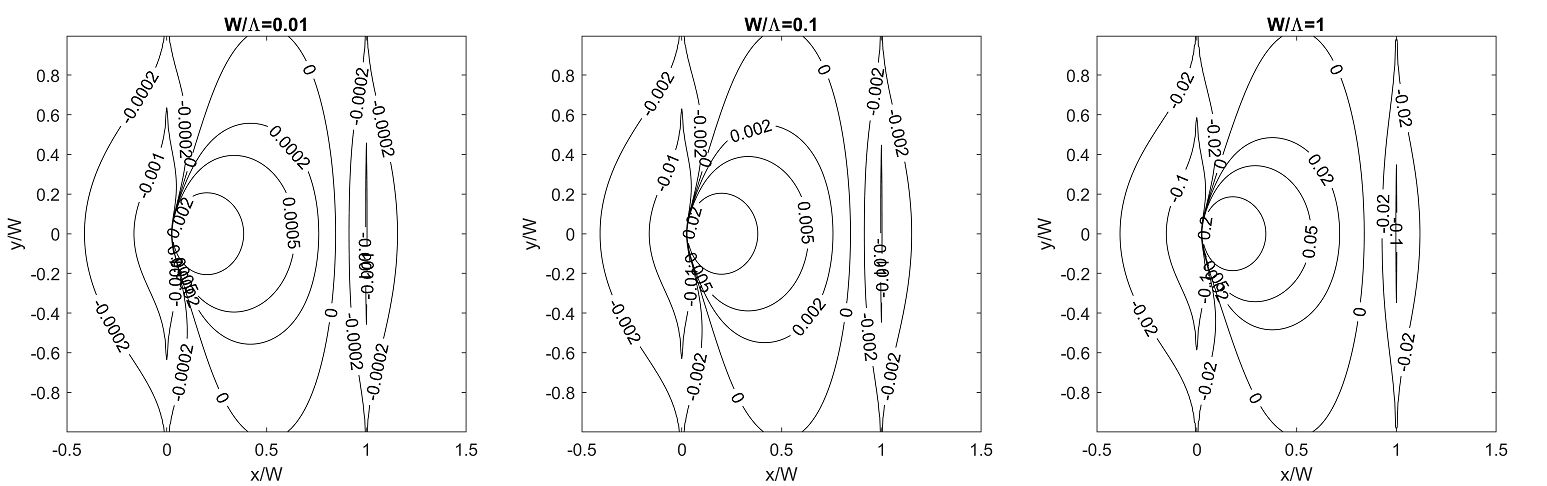}
	\caption{Contours of the field $h_z(x,y)=$ constant in  units of $\phi_0/W^2$ for varying strip widths $W/\Lambda=\{0.01, 0.1, 1\}$  ($z=0$)   and the same  vortex position $a/W=0.1$.}
	\label{f5}
\end{figure}

  Fig.\,\ref{f6} exhibits a qualitative difference between the strip and the infinite film.  The comparison between the two cases shows the origin of the difference, i.e. the presence and absence of the flux return paths, or in our plots at $z=0$ the clear evidence of the change of $h_z$ sign.  Although the wide strip and  the infinite film  have nearly identical near-the-core field behavior, the field patterns differ  globally,    the strip field is stretched  in the strip direction and exhibits the strong edge effects.  
 
\begin{figure}[htbp]
	\centering
		\includegraphics[width=1.\textwidth]{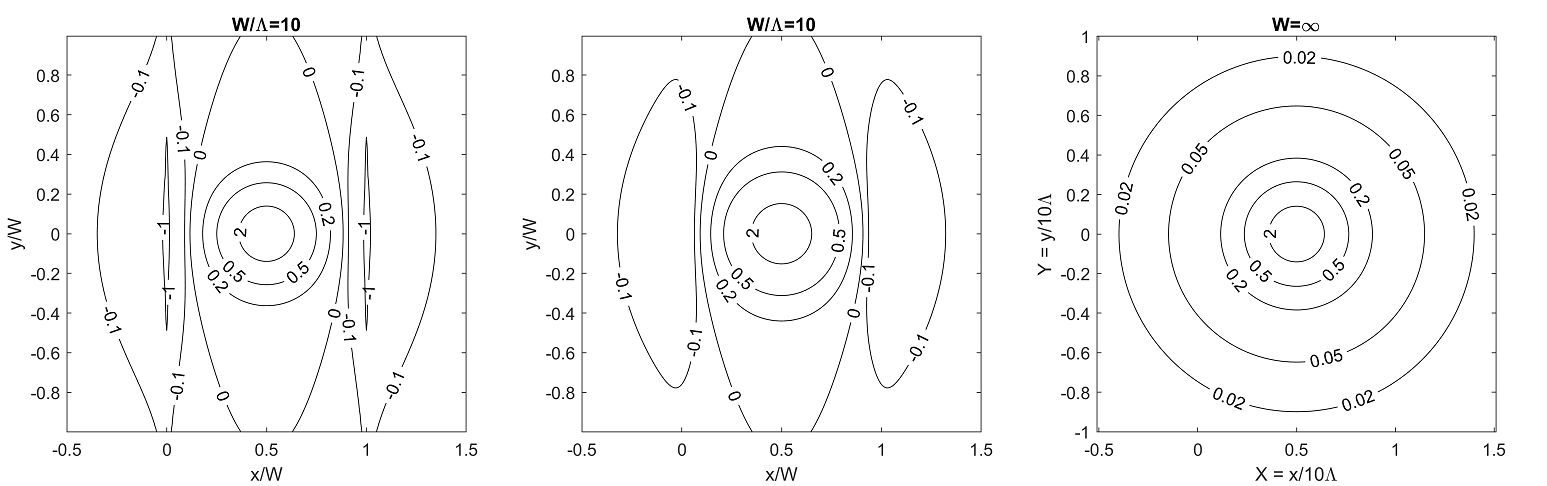}
	\caption{The left  panel: $h_z(x,y)$ of a wide strip ($W/\Lambda=10$) in  units of $\phi_0/W^2$ at the film surface $z=0$. The central panel: the same for    $z/W= 0.1 $. The right  panel is for   the infinite film, scaled and shifted to $a=0.5/10\,\Lambda$ for comparison.}
	\label{f6}
\end{figure}

The flux return affects   the integrated flux calculation.  As stated, the total flux through the entire $xy$ plane ($z$=const) is 0 for a strip, while it is $\phi_0$ for the infinite film.  The flux through the strip is shown in Fig. \ref{f7} versus vortex position.  The total flux through the strip surface is well below $\phi_0$ in particular due to the negative $h_z$ regions at the strip surface.  The positive flux, i.e. the restricted integral over the sub-region of the $xy$ plane ($z$=const) where $h_z\geq 0$, is greater than the total flux, though still short of $\phi_0$.  
Unlike the infinite film, the positive flux decreases when $z>0$ increases, and in fact the lift-off effect is substantial.

 \end{widetext}

\begin{figure}[htb ]
	\centering
		\includegraphics[width=0.35\textwidth]{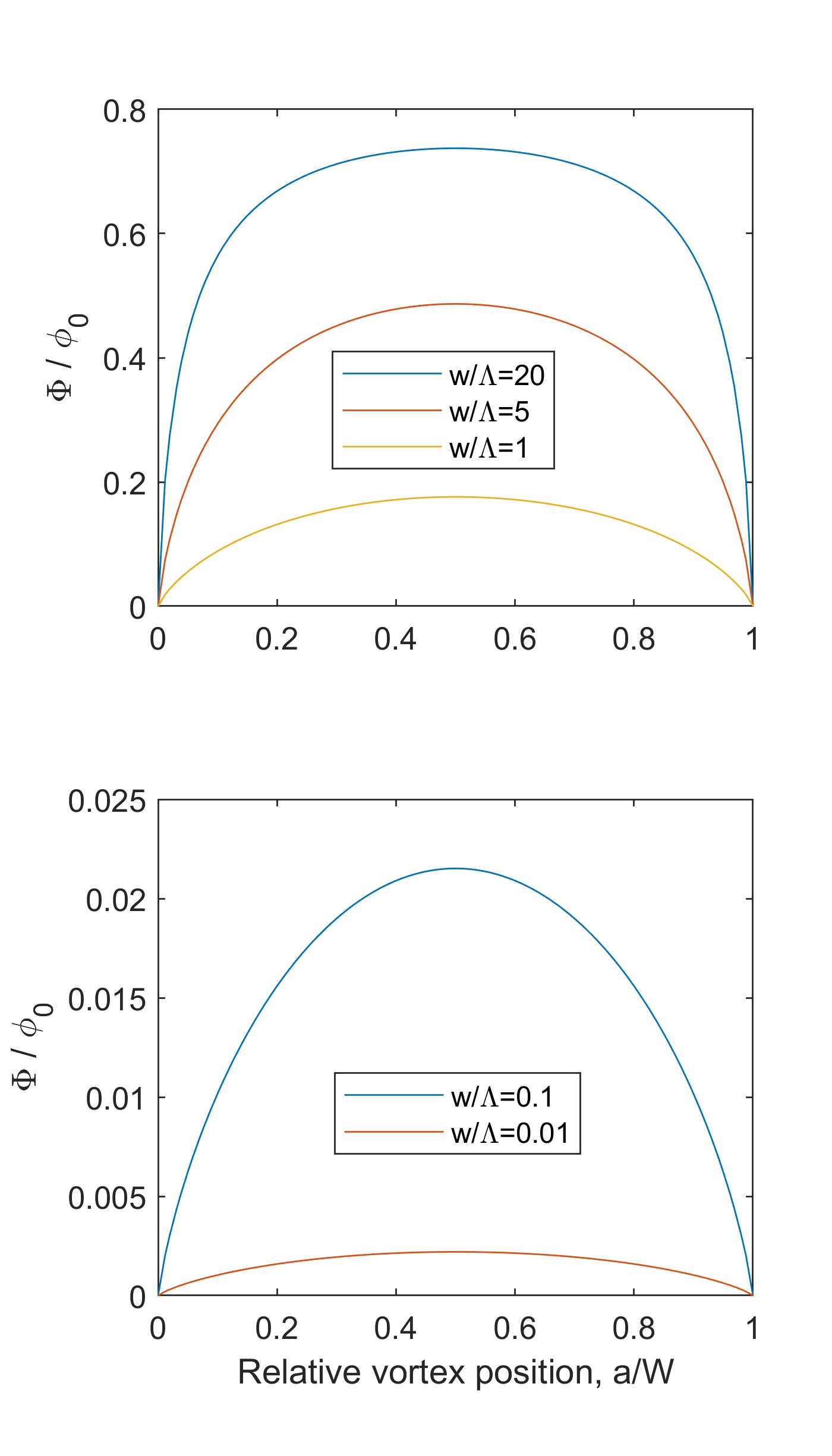}
	\caption{The magnetic flux through the strip surface  in the unit of $\phi_0$ as a function of the vortex position. The upper panel is for wide strips  $W/\Lambda=\{1,5,20\}$, the lower panel is  for narrow  ones, $W/\Lambda=\{0.01,0.1\}$.  }
	\label{f7}
\end{figure}

The energy (magnetic plus kinetic) is obtained from the general result for thin films $\epsilon=\phi_0S[\bm r\to(\bm a+\bm \xi)]/2c$ where $\bm \xi$ is a vector of the size of the vortex core (the coherence length) that should be added because $S(\bm a)$ diverges \cite{K94}, excluding also  small regions of the strip near   the edges.  The results for $\epsilon(\bm a)$ are shown in Fig. \ref{f8} for a set of strip widths  choosing $\Lambda/\xi=1000$.  The wide-strip energy in the strip middle approaches the value in the infinite film, albeit slowly.  
 
As stated, the common trend has been observed in  the field patterns which are nearly independent of the width for $W \alt \Lambda$.  The similarity may be viewed in their strength as well.  To this end, consider  dimensionaless quantities $\tilde{h}_z$ and $\tilde{S}$ 
\begin{equation}
	h_z = (\phi_0/W^2) \tilde{h}_z,\ S = (c\phi_0/2\pi\Lambda) \tilde{S}.
	\label{eNN4}
\end{equation}
In general, they are functions of $W/\Lambda$, in addition to $\{ 
\bm{r}/W , a/W \}$. However, the numerical results show that their magnitudes appear to exhibit simple trends, namely $\tilde{h}_z$ is roughly proportional to $W/\Lambda$, while $\tilde{S}$  is mostly width-independent, for given $a/W$ and in the said width range. Such approximate trends could  be useful in analyzing experimental data with samples of various widths.

%%%%%%%
\section{Discussion}
%%%%%%%

In this work we   study the currents and field distributions of  Pearl vortices at  thin-film superconducting strips. 
 The new feature which - to our knowledge - was not discussed in literature is that the vortex magnetic field lines crossing the film from the lower half-space ($z<0$ under the film) to the upper one ($z>0$) return back not only via a free space out of the strip edges, but also through some patches of the film {\it per se}, see examples in Figs.\,\ref{f3} and  \ref{f6}. As far as we know, these features of the field distribution have never been experimentally observed. However, with improved techniques of measuring {\it local} fields such as nitrogen-vacancy  centers in diamonds sensitive to the magnetic field, these observations become possible \cite{Kobayashi}. 
 
 Also, we evaluate the magnetic flux through the strip carried by a vortex and its energy as  functions of its position and their dependencies on the strip width $W$ and the Pearl length $\Lambda$ of the film. The vortex energy is relevant for the potential barrier the vortex should overcome while crossing the strip, the thermally activated or   the quantum tunneling processes \cite{Tafuri}.

The field distribution of vortices in films has been studied before, see, e.g., the work of 
H.\,Brandt \cite{Brandt} and references therein. This was done, however, for thick films, without going to the thin film limit, which has many peculiar features (as this work shows) and which is relevant for applications in superconducting electronics.

 \begin{figure}[t]
	 \centering
		 \includegraphics[width=0.4\textwidth]{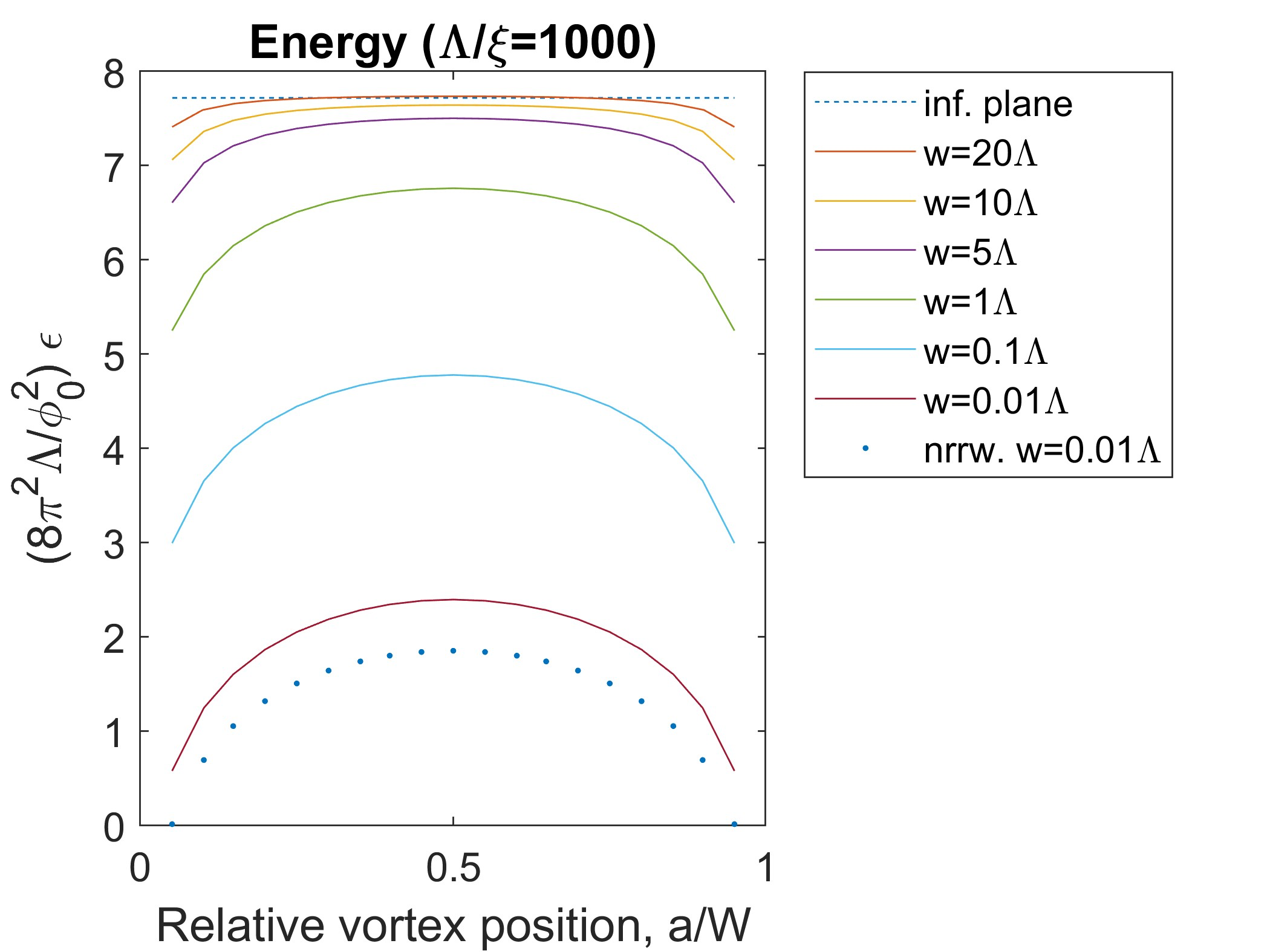}
	 \caption{The vortex energy versus its position    for set of widths given in the legend.  
The top dotted line is for an infinite film, the bottom one is for narrow limit according to Eq.\,(\ref{energy}).}
	 \label{f8}
 \end{figure}

%%%%%%%%
\appendix
%%%%%%%
 
\section{Outline of numerical procedure }

To obtain an equation for $S$, we rewrite Eq.\,(\ref{e2})   as
\begin{equation}
	h_z(\bm{r})c = \int d^2 \bm{r}' \bm{R} \cdot \bm{\nabla} S(\bm{r}') / R^3,
	\label{eNN5}
\end{equation}
where $\bm{R} = \bm{r} - \bm{r}'$ and  the integral is over the strip surface.  Elimination of $h_z$ from Eqs.~(\ref{eNN3}) and (\ref{eNN5}) yields
\begin{align}
S(\bm{r}) + \frac{1}{2\pi\Lambda} & \int d^2 \bm{r}' \overline{G} (\bm{r},\bm{r}') \int d^2 \bm{r}'' 
{\bm{R} \cdot \bm \nabla S(\bm{r}'') / R^3} \rvert _{\bm {R=r'-r''}} \nonumber
\\
&= (c\phi_0/2\pi\Lambda) \overline{G} (\bm{r},\bm{a})\,.
	\label{eNN6}
\end{align}
Once $S$ is obtained, the field $h_z(\bm r)$ can be calculated with the help of Eq.\,(\ref{eNN5}).

In principle, Eq.\,(\ref{eNN6}) can be discretized by expansion in $\chi$ where $\overline{G}$ is diagonal, while the remaining kernel has off-diagonal elements of the form
\begin{equation}
	L_{mn}^{k_y} \equiv \int d^2 \bm r \int d^2 \bm r' { \chi _m^{k_y} (\bm r) \bm{R \cdot \nabla} \chi_n^{k_y} (\bm r') ^\ast } /R^3.
	\label{eNN8}
\end{equation}
In practice, however,   it is necessary to isolate the divergent
behavior of $S$ at the vortex position, otherwise the
expansion will have the ``ringing" problem (Gibbs phenomenon).
To this end,  we extract the divergent part of $S$  analytically.  Explicitly, we write $S=\hat{S}+\Delta S$ with  
\begin{align}
	\hat{S} (\bm r) = -\frac{c\phi_0}{8\pi^2\Lambda} 
	\{ &\ln [ (x-a)^2 + y^2 ] - (1-x) \ln (a^2+y^2) \nonumber
	\\ &- x \ln ((1-a)^2+y^2) \}
	\label{eNN9}
\end{align} 	 
where, in the curly brackets,  all lengths are in units of $W$.  The remaining $\Delta S$ is finite and continuous everywhere on the strip, and thus amenable to the expansion with a finite number of terms.

It should be noted that the expansion in the $y$ direction amounts to the Fourier transform, meaning that the numerical calculation of $\Delta S$ is carried out in the $k_y$ Fourier space, and the results are converted to the $y$ space by the fast Fourier transform.  The evaluation of Eq.\,(\ref{eNN8}) can be simplified by utilizing the representation
	\begin{equation}
	\int \frac{dy\,e^{i k_y y}}{R} = 2K_0(|k_y x|) = \int_0^{\infty} \frac{du}{ u } e^{-(u+u^{-1})|k_y x|/2}.\qquad
\end{equation}

\end{document}